\documentclass[runningheads]{llncs}
\usepackage[T1]{fontenc}
\usepackage{xcolor}
\usepackage{graphicx}
\usepackage{amsmath}
\usepackage{amssymb}
\usepackage{booktabs}
\usepackage{multirow}
\usepackage{tabularx}
\usepackage{tabularray}
\usepackage{svg}
\usepackage{subfig}
\usepackage{tabularray}

\begin{document}
\title{Robust Federated Anomaly Detection Using Dual-Signal Autoencoders: Application to Kidney Stone Identification in Ureteroscopy}
\titlerunning{Robust Federated Detection for Kidney Stone}
%
\author{Ivan Reyes-Amezcua\inst{1} \and Francisco Lopez-Tiro\inst{2,3} \and Clément Larose\inst{4} \and  \\ Christian Daul\inst{3} \and   Andres Mendez-Vazquez\inst{1} \and Gilberto Ochoa-Ruiz\inst{2}}

\authorrunning{I. Reyes-Amezcua et al.}
%
\institute{CINVESTAV, Guadalajara, Mexico \and
Tecnologico de Monterrey, School of Engineering and Sciences, Mexico \and CRAN UMR 7039, Université de Lorraine and CNRS, Nancy, France
\and CHRU de Nancy-Brabois, Service d'urologie, Vandœuvre-les-Nancy, France  
}
%
\maketitle              
\begin{abstract}
This work introduces Federated Adaptive Gain via Dual Signal Trust (FedAgain), a novel federated learning algorithm designed to enhance anomaly detection in medical imaging under decentralized and heterogeneous conditions. Focusing on the task of kidney stone classification, FedAgain addresses the common challenge of corrupted or low-quality client data in real-world clinical environments by implementing a dual-signal trust mechanism based on reconstruction error and model divergence. This mechanism enables the central server to dynamically down-weight updates from untrustworthy clients without accessing their raw data, thereby preserving both model integrity and data privacy. FedAgain employs deep convolutional autoencoders trained in two diverse kidney stone datasets and is evaluated in 16 types of endoscopy-specific corruption at five severity levels. Extensive experiments demonstrate that FedAgain effectively suppresses "expert forger" clients, enhances robustness to image corruptions, and offers a privacy-preserving solution for collaborative medical anomaly detection. Compared to traditional FedAvg, FedAgain achieves clear improvements in all 16 types of corruption, with precision gains of up to $+14.49\%$ and F1 score improvements of up to $+10.20\%$, highlighting its robustness and effectiveness in challenging imaging scenarios.

\end{abstract}

\section{Introduction}

Deep learning has significantly advanced medical image analysis, notably enhancing tasks such as kidney stone classification \cite{bi2019artificial,hamdi2022evaluation}. However, conventional methods typically rely on local training approaches, limiting data diversity due to the process and/or instrument for the acquisition, and exposing models to performance degradation under perturbations like noise, blur, illumination changes, and camera instability. Such perturbations, collectively termed image corruptions, frequently arise in clinical environments, underscoring the necessity for robust and reliable models that ensure diagnostic certainty in medical practice \cite{kaissis2020secure}.

Federated Learning (FL) provides a strategic response to these limitations by enabling multiple medical centers to collaboratively train models without directly sharing sensitive patient data \cite{ng2021federated}. Each client independently trains a local model and communicates only model parameters to a central server for aggregation, effectively addressing privacy concerns while incorporating diverse and decentralized data \cite{reyes2024leveraging,laridi2024enhanced}. Nevertheless, federated training is vulnerable to corrupted datasets affected by artifacts like blur, noise, or poor lighting, which can degrade global model performance.

To effectively identify and mitigate these image corruptions in a federated, multiclient environment, autoencoder-based anomaly detection methods have emerged as promising solutions. Autoencoders detect deviations using reconstruction errors, successfully identifying medical anomalies such as tumors and imaging artifacts \cite{laridi2024enhanced,cai2024rethinking}. Unlike standard convolutional neural networks (CNNs), autoencoders excel by learning intrinsic data representations, making them particularly effective at capturing subtle anomalies. Integrating autoencoders into federated learning frameworks enables the collaborative, privacy-preserving detection of anomalies, significantly improving the robustness and generalization of models across varied clinical settings \cite{shvetsova2021anomaly}.

\begin{figure*}[!t]
\centering
\includegraphics[width=\linewidth]{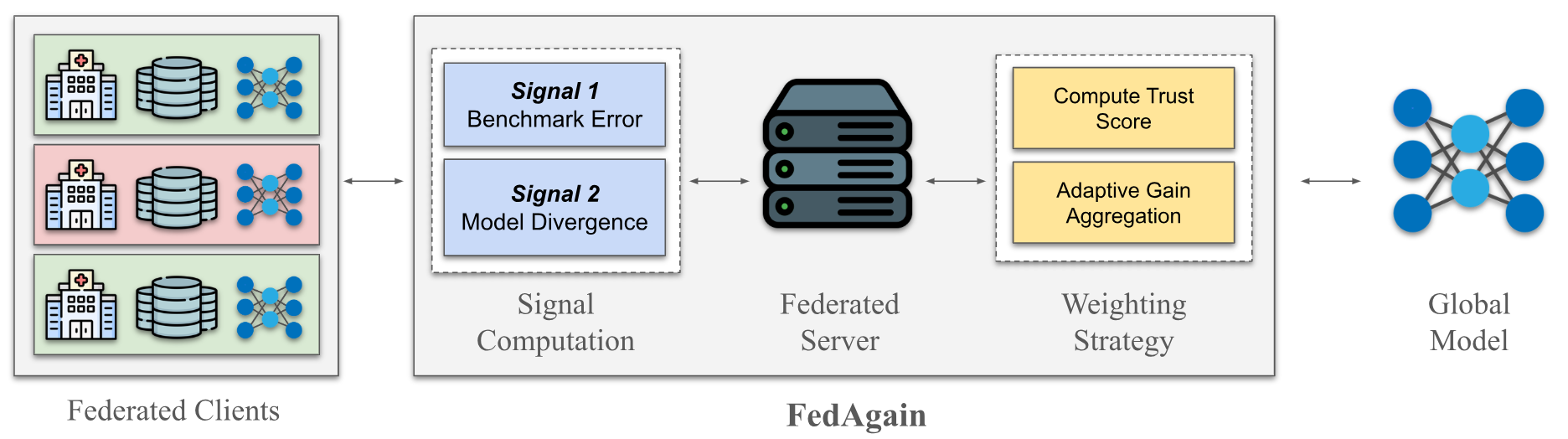}
\vspace{-5mm}
\caption{Overview of the FedAgain framework. Federated clients (left) train local models and send updates to a central server. Clean clients are shown in green, while corrupted clients are highlighted in red. The server computes two trust signals: benchmark error and model divergence, which are used to assign trust scores. These scores guide the adaptive gain aggregation process, reducing the impact of unreliable clients. The aggregated result is used to update the global model.
}
\label{fig:overview}
\end{figure*}

Federated autoencoder-based anomaly detection is still underexplored in kidney stone imaging, mainly due to limited annotated data. This study addresses the gap using a specialized dataset with endoscopic images and synthetic corruptions, enabling systematic evaluation of model robustness in federated settings.
A key challenge addressed in this work is the \textit{expert forger} problem, which arises when a client consistently submits corrupted data (e.g., from persistently blurry endoscopic cameras), inadvertently training its local model to specialize in these corruptions. This specialization compromises the generalizability of the global model when aggregated via conventional Federated Averaging (FedAvg) \cite{mcmahan2017communication}. To mitigate this issue, a novel federated autoencoder framework, \textit{FedAgain} (Federated Adaptive Gain via Dual Signal Trust) is introduced. FedAgain incorporates a dual-signal trust mechanism to dynamically evaluate and down-weight untrustworthy clients, ensuring enhanced resilience, robustness, ultimately, greater diagnostic certainty (Figure \ref{fig:overview}).

This paper introduces the first federated autoencoder framework specifically tailored for robust endoscopic kidney stone imaging. Using six specialized autoencoder models, the approach is rigorously evaluated across 16 different corruption types, employing various performance metrics to ensure comprehensive analysis.

The rest of the paper is organized as follows: Section 2 details the methodology, including the FL framework, autoencoder architecture, and the trust mechanism employed by FedAgain. Section 3 describes the datasets and preprocessing. Section 4 covers corruption simulation and evaluation metrics. Section 5 presents results and client behavior analysis. Section 6 concludes with key findings and implications.

\section{Related Work}

FL first gained traction with FedAvg, which averages local updates instead of sharing raw data, but it falters when even a few clients are corrupted \cite{mcmahan2017communication}.  Aggregation-based defenses such as FLTrust attach every round to a small, server-held “root” dataset, sharply curbing Byzantine influence \cite{cao2020fltrust}, while Bayesian ensembling in FedBE samples and averages multiple global models to stay stable on highly non-IID (independent and identically distributed) data \cite{chen2020fedbe}. However, a recent systematic survey notes that these strategies still leave long-term, client-specific artifacts largely unaddressed \cite{uddin2025systematic,guan2024federated}. Autoencoder anomaly detectors fill part of this gap: autoencoders minimize bandwidth and spot outliers at the edge \cite{novoa2023fast}, and global summary-statistics thresholds further boost detection under skewed client distributions \cite{laridi2024enhanced}.  Nevertheless, most autoencoder work targets generic noise or blur and rarely tackles the persistent, domain-specific corruptions common in clinical images, an issue repeatedly highlighted in medical-imaging FL reviews \cite{kaissis2020secure,guan2024federated}.

Kidney stone ureteroscopic data highlight challenges such as glare, blur, and color drift across hospitals \cite{lopez2021assessing,lopez2024vivo}. A corruption-aware FL system combines a pre-trained CNN with two stages: Learning Parameter Optimization and Federated Robustness Validation on corrupted data \cite{reyes2024leveraging}. These results support FL pipelines that integrate robust aggregation and client-side anomaly detection for reliable, privacy-preserving diagnosis.



\section{Federated Adaptive Gain via Dual-Signal Trust}


FL allows clients (e.g., hospitals) to collaboratively train a shared model without sharing raw data, preserving privacy and ensuring compliance (e.g., HIPAA, GDPR). Clients send only model updates, reducing communication costs and enabling scalable, privacy-preserving training on distributed medical data.

FL offers a principled solution by training models where the data reside, but standard FL algorithms remain vulnerable to persistent data-quality heterogeneity: A handful of \textit{noisy} or corrupted clients can bias model updates and degrade performance \cite{li2020federated}. The problem is especially pronounced in kidney stone imaging, where differences in endoscopic and laboratory acquisition pipelines routinely produce specular reflections, blur, sensor noise, and lens artifacts that vary from one clinical site to another. This leads to a critical vulnerability referred to as the \textit{expert forger} problem. A client with a persistent, local data corruption (e.g., a perpetually blurry camera) will train its local model to become an \textit{expert} at reconstructing that specific corruption. In conventional FedAvg, the parameter updates from such an \textit{expert-forger} client are incorporated into the global weight average, contaminating the aggregated model and substantially degrading its ability to recognize the corresponding defect as an anomaly.

FedAgain is a strategy that builds robust anomaly detection models by weighting client contributions based on trust scores, replacing uniform averaging with a two-phase process: client-side dual-signal generation (reconstruction error and model divergence), followed by server-side trust assessment and adaptive aggregation, see Figure \ref{fig:overview}.

\subsection{FedAgain Algorithm}
Each step in FedAgain comprises two phases: A client one (Dual-Signal generation) followed by a server phase (Trust assessment and adaptive aggregation), see Figure \ref{fig:overview}.

\subsubsection{Client phase:}
For every selected client $C_k$:

\begin{enumerate}
    \item \textit{Benchmark signal:} It evaluates  incoming global models $w_t$ on the local dataset $\mathcal{D}_k$ by using:
    \begin{equation}
        E_k = \frac{1}{|\mathcal{D}_k|}\sum_{(x_i,y_i)\in \mathcal{D}_k}
                 \ell\!\bigl(f_{w_t}(x_i),\,y_i\bigr).
    \end{equation}
    where $\ell$ is a loss function that measures the discrepancy between the prediction of the model $f_{w_t}(x_i)$ and the true label $y_i$. The resulting scalar $E_k$ is sent back to the server. A high value of $E_k$ indicates that the client’s data distribution $\mathcal{D}_k$ differs significantly from the global distribution.

    \item \textit{Local adaptation:} It performs $e$ epochs of stochastic gradient descent on $\mathcal{D}_k$ to obtain an updated model $w_{t+1}^{(k)}$. Finally, it transmits $w_{t+1}^{(k)}$ to the server.

\end{enumerate}

\subsubsection{Server phase:} For each responding client:

\begin{enumerate}
    \item \textit{Divergence signal: } It measures the results with the Euclidean or cosine distance by defining 
    \begin{equation}\label{eq:divergence}
        D_k = \bigl\|\,w_{t+1}^{(k)} - w_t\bigr\|,
    \end{equation}
    A large $D_k$ implies that substantial parameter changes are required to fit the local data.

    \item \textit{Dual-signal trust score:} It is calculated with a small constant $\varepsilon$ to ensure numerical stability by
    \begin{equation}\label{eq:trust}
        T_k \;=\; \frac{1}{E_k\,D_k + \varepsilon}.
    \end{equation}
    The score sharply discounts clients exhibiting high error, high divergence, or both.

    \item \textit{Adaptive gain aggregation: } The next global model is the trust-weighted average (Eq. \ref{eq:trust-weight})

    \begin{equation}\label{eq:trust-weight}
        w_{t+1} \;=\; \frac{\displaystyle\sum_{k} T_k\,w_{t+1}^{(k)}}
                         {\displaystyle\sum_{k} T_k}.
    \end{equation}
\end{enumerate}

Clients with high error and divergence receive low aggregation weight, while reliable clients contribute more. Repeated trust-weighted updates yield a robust, privacy-preserving global model.

\subsection{Autoencoder Architecture}

A deep convolutional autoencoder, trained on clean data, encodes kidney stone images into a 128-dimensional latent space and detects anomalies via reconstruction errors. It serves as both the global model and client evaluator, with the server distributing the current global autoencoder $w_t$ each round to represent the clean image distribution. Each client $C_k$ first passes its local dataset through $w_t$ to obtain a mean-squared reconstruction error $E_k$ (\textbf{\textit{Signal 1}}). Next, the client then fine-tunes the autoencoder on its data, producing an updated model $w_{t+1}^{(k)}$. The server measures the parameter divergence (Eq. \ref{eq:divergence}) (\textbf{\textit{Signal 2}}) and derives a trust weight (Eq. \ref{eq:trust}). These weights are used to compute the next global model via a trust-weighted average (Eq. \ref{eq:trust-weight}).

Thus, the autoencoder functions simultaneously as benchmarking tool, trainable local model, divergence source, and aggregation target, enabling FedAgain to down-weight corrupted clients and iteratively refine a robust, privacy-preserving global detector.

\section{Datasets and Case Study}\label{sec:datasets}




Kidney stone formation is a significant public health concern, especially in industrialized countries, affecting up to 10\% of the population and showing recurrence rates of approximately 40\% in the United States. Therefore, accurate identification of stone types is crucial for personalized treatment and effective prevention strategies \cite{corrales2021classification}. The current standard for kidney stone classification, Morpho-Constitutional Analysis (MCA), integrates visual examination and biochemical evaluation. However, MCA is time-consuming, reliant on expert interpretation, and unsuitable for real-time clinical use. An alternative, Endoscopic Stone Recognition (ESR), performed during ureteroscopy, remains subjective and is further complicated by variable lighting conditions, motion artifacts, and image noise \cite{el2022evaluation}.

To address these challenges, deep learning models have emerged as a promising solution to automate stone classification from endoscopic and non-endoscopic images \cite{el2022evaluation,lopez2024vivo}. However, existing models are typically trained on local datasets, each presenting distinct challenges for classification and segmentation tasks. Consequently, there is a critical need for robust models capable of maintaining consistent performance in the presence of anomalies and artifacts, while also facilitating effective collaboration between multiple institutions.

Integrating this task within a FL framework offers a practical solution, enabling multiple institutions to collaboratively train models without sharing sensitive image data directly. Furthermore, FL provides an effective platform to assess and enhance model robustness against the variety of real-world artifacts encountered during clinical practice.

\subsection{Datasets}

In this contribution, two ex‑vivo (extracted) kidney stone datasets were used for the evaluation of FedAgain. The key characteristics of these datasets are detailed as follows:

Dataset A \cite{corrales2021classification} includes 366 CCD images (Charge-Coupled Device) \textcolor{blue} of kidney stone fragments. Dataset A images were acquired under controlled illumination conditions, minimizing reflections and camera motion, with high spatial resolution  (4288$\times$2848 pixels) and image quality, and a uniform background.

Dataset B \cite{el2022evaluation} contains 409 endoscopic images acquired with a flexible ureteroscope in an environment that simulates real acquisition conditions. Dataset B is close  in a simulated clinical environment. Images in dataset B are visually similar to those captured during in-vivo utereroscopy, including changes in lighting, blur caused by camera movement, and limited spatial resolution (1920$\times$1080 pixels).

All images were manually segmented under expert supervision. Preprocessing follows the protocol of \cite{lopez2021assessing}, ensuring normalization and balancing of stone classes while excluding tissue background. Stone patches of fixed size 256×256 pixels are extracted to support robust training and evaluation splits of 80\% for training and 20\% for testing, with strict measures to prevent data leakage \cite{lopez2024vivo}.



\begin{figure*}[!h]
\centering
\includegraphics[width=0.95\linewidth]{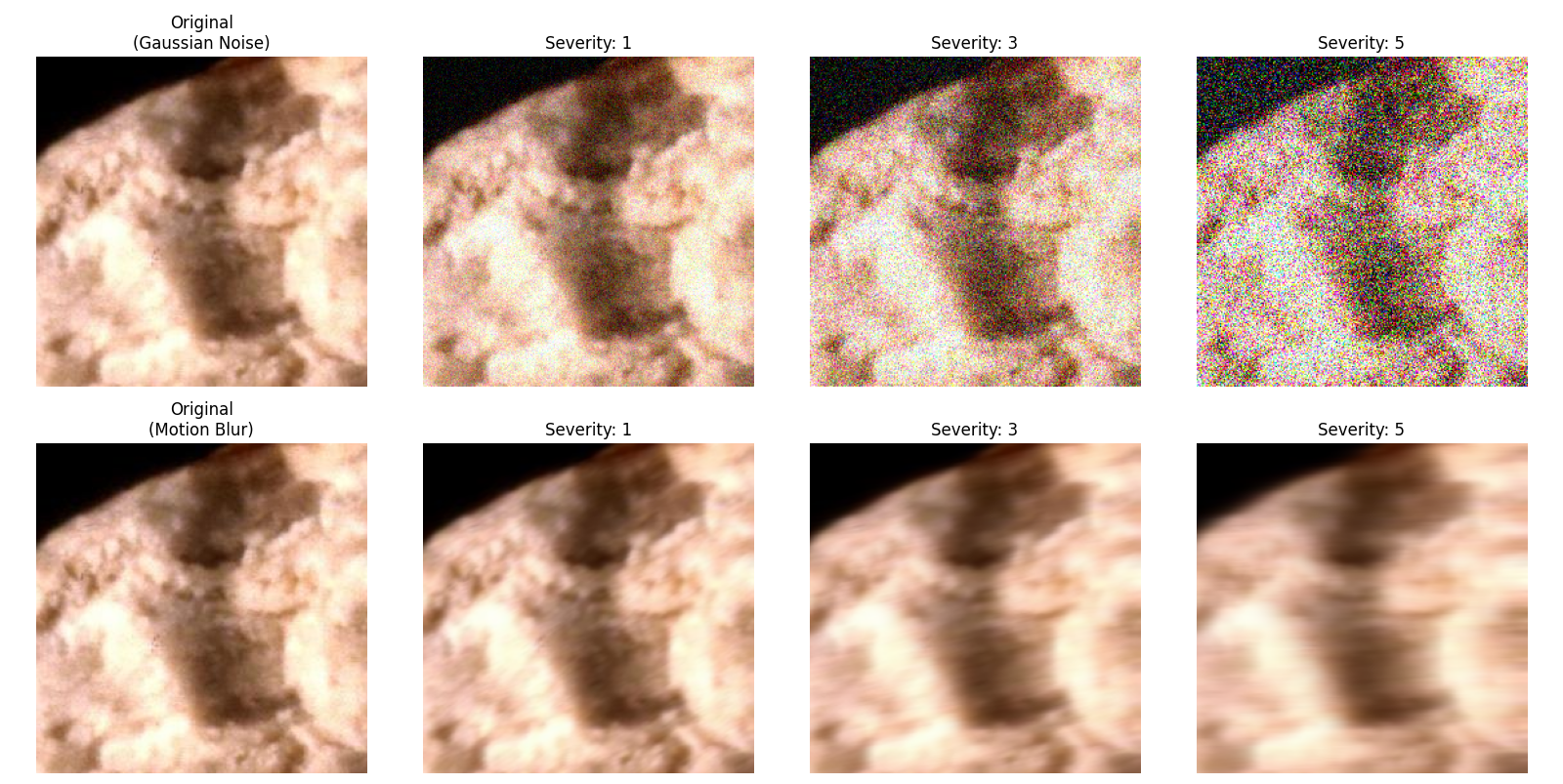}
\vspace{-3mm}
\caption{Visualization of four representative image corruptions applied to a kidney stone endoscopic image. Each row shows the original image followed by increasing levels of corruption severity (1, 3, and 5).}
\label{fig:corruptions}
\end{figure*}

\subsection{Endoscopy Corruptions}
\label{sec:endoscopy_corruptions}

In this work, the existing framework provided by the \textit{endoscopy corruptions} \cite{reyes2024endodepth} library is leveraged to simulate clinically realistic image degradations in the kidney stone datasets. This library offers a comprehensive suite of 16 corruption functions, each designed to replicate visual artifacts commonly observed in endoscopic imaging, such as specular reflections, motion blur, sensor noise, and lens distortions.

Each corruption function accepts a clean input image and a severity level (1 to 5), see Figure \ref{fig:corruptions}, enabling a fine-grained evaluation of model robustness. By applying these perturbations to the dataset used in this study, a controlled experimental setup is created to assess the sensitivity of the proposed federated autoencoder to diverse and increasingly severe real-world image corruptions.

\subsection{Configuration Setup}

A federated learning setup with 20 clients is simulated, selecting 10 per round over 20 rounds. Each client trains a convolutional autoencoder locally for 3 epochs. While class labels are IID, 30\% of clients receive uniquely corrupted images, introducing non-IID data quality. The global model is evaluated on 16 corruption types at 5 severity levels using the \textit{endoscopycorruptions} library, with performance assessed by classification metrics.

\begin{table}[!t]
\centering
\caption{Accuracy performance across 16 corruption types at severity level 4 for different dataset variants. Column headers follow the format \texttt{<Dataset>-<View>}, where \texttt{A} and \texttt{B} denote the datasets, and \texttt{MIX}, \texttt{SEC}, and \texttt{SUR} indicate mixed, section, and surface views, respectively. Highest values per row are in bold.}

\label{tab:corruption_acc}
\begin{tabular}{lcccccc}
\toprule
\textbf{Corruption} & \textbf{A-MIX} & \textbf{A-SEC} & \textbf{A-SUR} & \textbf{B-MIX} & \textbf{B-SEC} & \textbf{B-SUR} \\
\midrule
brightness           & 0.686 & \textbf{0.717} & 0.537 & 0.525 & 0.536 & 0.565 \\
color\_changes       & 0.615 & 0.645 & 0.565 & 0.707 & 0.718 & \textbf{0.750} \\
contrast             & 0.500 & 0.500 & \textbf{0.500} & 0.500 & 0.500 & 0.500 \\
darkness             & 0.500 & 0.500 & 0.500 & 0.516 & \textbf{0.528} & 0.503 \\
defocus\_blur        & \textbf{0.501} & 0.500 & 0.500 & 0.500 & 0.500 & 0.500 \\
fog                  & 0.648 & 0.698 & 0.574 & 0.679 & \textbf{0.728} & 0.719 \\
gaussian\_noise      & 0.972 & 0.968 & 0.986 & \textbf{1.000} & \textbf{1.000} & \textbf{1.000} \\
glass\_blur          & 0.516 & 0.521 & 0.503 & 0.516 & \textbf{0.535} & 0.531 \\
impulse\_noise       & 0.967 & 0.987 & 0.981 & \textbf{1.000} & \textbf{1.000} & \textbf{1.000} \\
iso\_noise           & 0.958 & 0.968 & 0.973 & \textbf{1.000} & \textbf{1.000} & \textbf{1.000} \\
lens\_distortion     & 0.762 & 0.839 & 0.753 & 0.949 & \textbf{0.950} & 0.944 \\
motion\_blur         & 0.500 & 0.500 & \textbf{0.500} & 0.500 & 0.500 & 0.500 \\
resolution\_change   & 0.500 & 0.500 & \textbf{0.500} & 0.500 & 0.500 & 0.500 \\
shot\_noise          & 0.990 & 0.987 & 0.991 & 0.999 & \textbf{1.000} & 0.999 \\
specular\_reflection & 0.670 & 0.735 & 0.648 & \textbf{0.828} & 0.804 & 0.820 \\
zoom\_blur           & 0.500 & 0.500 & \textbf{0.500} & 0.500 & 0.500 & 0.500 \\
\bottomrule
\end{tabular}
\end{table}

\section{Experiments and Evaluation}
The proposed method was evaluated on two kidney stone datasets with various realistic corruptions, revealing its ability to handle decentralized, heterogeneous data and distinguish between trustworthy and unreliable clients.
Table \ref{tab:corruption_acc} presents the classification accuracy for 16 corruption types at severity level 4, evaluated across datasets A and B and three imaging views: mixed, section, and surface. The results reveal that both the dataset domain and the imaging perspective influence how well the model handles degradation. Some corruptions, such as Gaussian and ISO noise, are consistently easier to detect, while others like resolution loss prove more challenging.

\begin{table}[!b]
\centering
\caption{Average percentage improvement over FedAvg \cite{mcmahan2017communication} across multiple metrics and dataset variants. Highest values per row are highlighted in bold.}
\label{tab:fedavg_vs_fedagain}
\begin{tabular}{lcccccc}
\toprule
\textbf{Metric} & \textbf{A-MIX} & \textbf{A-SEC} & \textbf{A-SUR} & \textbf{B-MIX} & \textbf{B-SEC} & \textbf{B-SUR} \\
\midrule
AUC       & \textbf{2.66\%} & 1.77\% & 0.13\% & 0.89\% & 1.34\% & 0.62\% \\
Accuracy  & 0.80\% & 0.65\% & 0.13\% & 0.39\% & \textbf{0.98\%} & 0.48\% \\
Recall    & \textbf{7.78\%} & 3.90\% & 2.06\% & 0.56\% & 6.12\% & -0.27\% \\
Precision & \textbf{14.49\%} & 3.45\% & 4.20\% & -2.85\% & 5.99\% & -2.05\% \\
F1        & \textbf{10.20\%} & 3.65\% & 2.74\% & -0.72\% & 5.31\% & -0.42\% \\
\bottomrule
\end{tabular}
\end{table}

Performance gains become clearer when comparing against the standard FedAvg approach. As shown in Table \ref{tab:fedavg_vs_fedagain}, the A-MIX configuration of FedAgain achieves substantial improvements, with precision rising by up to $+14.49\%$, recall by $+7.78\%$, and F1 score by $+10.20\%$. These gains reflect stronger anomaly detection and fewer false positives. Configurations A-SEC and A-SUR also show moderate benefits, while B-MIX and B-SUR yield smaller or even negative changes. Overall, the results point to the advantage of FedAgain, particularly when trained on diverse data distributions.

\begin{figure*}[!t]
\centering
\includegraphics[width=\linewidth]{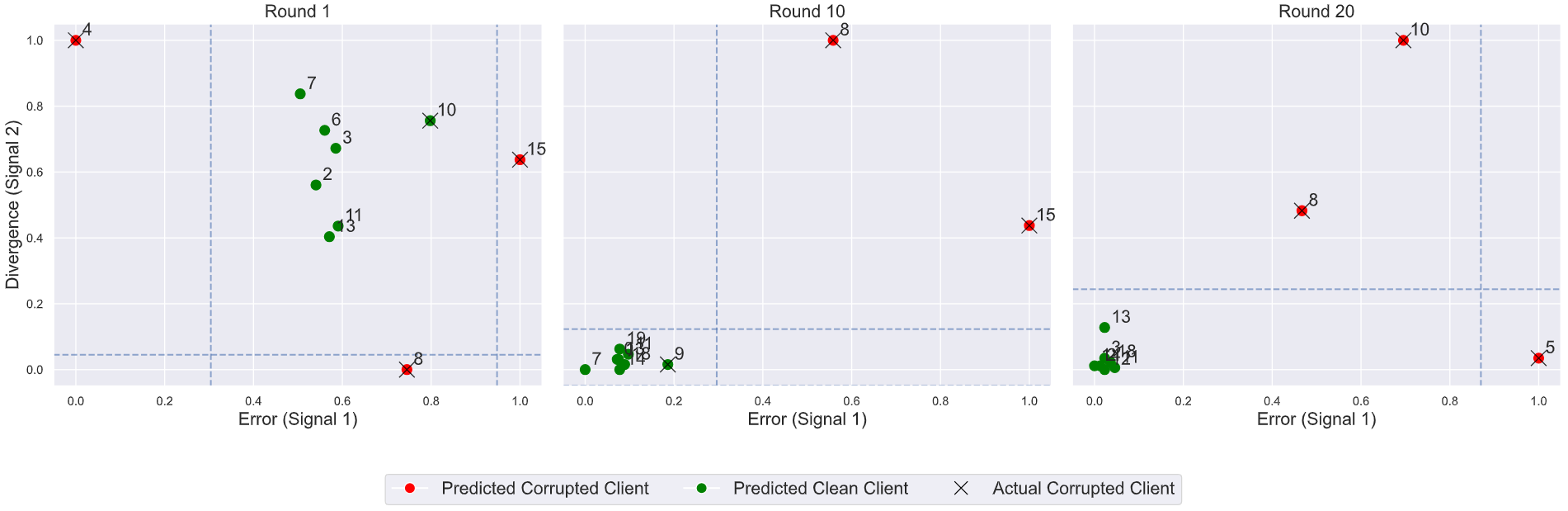}
\vspace{-5mm}
\caption{Visualization of client behavior at rounds 1, 10, and 20 using FedAgain on Dataset A-SEC. The plot shows increasing separation between trustworthy and untrustworthy clients over time.}
\label{fig:identification}
\end{figure*}


Figure \ref{fig:identification} illustrates the evolution of client trust across rounds 1, 10, and 20. Initially, clean and corrupted clients are indistinguishable. By round 10, clean clients cluster in the low-error, low-divergence region, while corrupted ones begin to drift. By round 20, separation is clear, showing how FedAgain progressively identifies and down-weights unreliable clients, enhancing robustness and preserving privacy.

\section{Conclusion}

This work proposes FedAgain, a federated learning framework for robust anomaly detection in kidney stone imaging. By combining reconstruction error and model divergence as trust signals, FedAgain mitigates the \textit{expert forger} problem caused by corrupted clients. Six specialized auto-encoders were trained in various configurations, and simulations with persistent corruptions showed effective anomaly detection and model robustness. FedAgain offers a scalable, privacy-preserving solution for medical imaging in federated settings.

\section*{Acknowledgments}
The authors wish to acknowledge the Mexican Secretaría de Ciencia, Humanidades, Tecnología e Innovación (Secihti) and CINVESTAV for their support in terms of postgraduate scholarships in this project, and the Data Science Hub at Tecnologico de Monterrey for their support on this project.
This work has been supported by Azure Sponsorship credits granted by Microsoft's AI for Good Research Lab through the AI for Health program.
The project was also supported by the French-Mexican ANUIES CONAHCYT Ecos Nord grant 322537.
We also gratefully acknowledge the support from the Google Explore Computer Science Research (CSR) Program for partially funding this project through the LATAM Undergraduate Research Program.

\bibliographystyle{splncs04}
\bibliography{biblio}

\end{document}